\documentstyle[amstex,amssymb,preprint,aps]{revtex}

\newtheorem{theorem}{Theorem}
\newtheorem{acknowledgement}[theorem]{Acknowledgement}

\begin{document}
\title{Local Relaxation and Collective Stochastic Dynamics }
\author{H. Neal Bertram and Xiaobin Wang}
\address{Center for Magnetic Recording Research, University of California,\\
San Diego, 9500 Gilman Drive, La Jolla, Ca \ 92093-0401}
\date{\today}
\maketitle

\begin{abstract}
Damping and thermal fluctuations have been introduced to collective normal
modes of a magnetic system in recent modeling of dynamic thermal
magnetization processes. The connection between this collective stochastic
dynamics and physical local relaxation processes is investigated here. A
system of two coupled magnetic grains embedded in two separate oscillating
thermal baths is analyzed with no {\it a priori} assumptions except that of
a Markovian process. It is shown explicitly that by eliminating the
oscillating thermal bath variables, collective stochastic dynamics occurs in
the normal modes of the magnetic system. The grain interactions cause local
relaxation to be felt by the collective system and the dynamic damping to
reflect the system symmetry. This form of stochastic dynamics is in contrast
to a common phenomenological approach where a thermal field is added
independently to the dynamic equations of each discretized cell or
interacting grain. The dependence of this collective stochastic dynamics on
the coupling strength of the magnetic grains and the relative local damping
is discussed.
\end{abstract}

\pacs{75.50.Tt, 75.40.Mg, 76.60.Es}

\section{Introduction}

A new model has recently been developed \cite{val0}, \cite{wang0}, \cite
{bert} to study thermal noise and dynamic thermal reversal in interacting
magnetic systems. In this approach, damping and thermal fluctuations are
introduced to the independent normal modes of the magnetic system,
corresponding to the analogy of temperature defined by independent particles
in an ideal gas. The damping term in the dynamic equations differ from that
of Landau-Lifshitz \cite{LLG} and, for even a single domain particle or
film, reflects the asymmetry of the magnetic energy \cite{bert}.
Generalization of the LLG equations to reflect the magnetic symmetry has
been discussed in general \cite{mills} with specific analysis for the
conduction electron relaxation process \cite{mills2}. Collective normal mode
processes have also been examined through analysis of relaxation to the
complete spin-wave spectrum in thin films \cite{michael}. Here we derive the
stochastic differential equations (SDE) specifically for the case of local
damping in a system of interacting grains.

Historically, stochastic differential equations have been developed by
simply adding a thermal fluctuation field to the Landau-Lifshitz-Gilbert
equations \cite{brown}. This approach has been widely utilized to analyze
the role of thermal fluctuations, for example, in non-uniformly magnetized
materials, such as a thin film, by discretizing the film and solving the
coupled LLG equations for each cell with a statistically independent random
field added to each cell \cite{zhu},\cite{borner}. One argument in favor of
this individual particle approach is that for physically localized
relaxation processes damping and thermal fluctuations can be conveniently
introduced to individual particles as an effective field. \ 

However, from a collective normal mode point of view, even physically
localized relaxation processes should give collective stochastic dynamics in
the normal modes. Application of this approach to analyze thermal noise in a
thin film appears to give better agreement with experiment than the
LLG-Brown approach \cite{jin}. Here, collective stochastic dynamics are
explicitly derived through a system-reservoir interaction model. We consider
two coupled magnetic grains embedded in two separate thermal baths, focusing
on small amplitude oscillations near equilibrium. No {\it a priori}
assumption is made concerning the form of the dynamic damping. This
configuration provides a simplified picture for localized relaxation
processes. We expand the analysis in \cite{bert} for a single grain
utilizing the method of \cite{scully} to add a generalized thermal bath to
the magnetization dynamics. \ The technique is to explicitly eliminate the
oscillating bath variables to obtain a closed stochastic equation for the
magnetic system. Under a Markovian approximation it can be shown that the
magnetic system obeys collective stochastic dynamics in the form of damped
harmonic oscillators in the normal modes. Thus, the original conjecture is
verified that damping and additive thermal fluctuations should be added to
the normal modes of a magnetic system, even if the physical relaxation
processes are local.

Section II introduces our model configuration. Section III obtains the
dynamic equations for the\ magnetic system by explicitly eliminating the
bath variables. In Section IV Markovian and rotational wave approximations
are utilized to obtain the collective stochastic dynamics for the
independent normal modes of a magnetic system. Section V discusses the
dependence of this collective stochastic dynamics on the magnetic
interaction and a comparison to the individual particle picture is given.

\section{Two Interacting Grains Embedded in Different Local Thermal Baths}

We consider two interacting nonidentical cubic magnetic grains of diameter $%
D $ embedded in two different localized thermal bathes, as shown in Fig.1.
Initially, neglecting the coupling to the thermal baths, the energy for the
host magnetic grains, normalized by $M_{s}^{2}V$, is:

\begin{eqnarray}
E_{m} &=&M_{s}^{2}V[\frac{1}{2}h_{k1}(1-m_{1z}^{2})+\frac{1}{2}%
h_{k2}(1-m_{2z}^{2})-h_{ex}\overrightarrow{m}_{1}\cdot \overrightarrow{m}_{2}
\label{ene} \\
&&+(\overrightarrow{m}_{1}\cdot \overrightarrow{m}_{2}-3m_{1x}m_{2x})] 
\nonumber
\end{eqnarray}
where $\overrightarrow{m}_{1}$ $=\overrightarrow{M}_{1}/M_{s}$and $%
\overrightarrow{m}_{2}=\overrightarrow{M}_{2}/M_{s}$ are normalized
magnetizations of the two grains. The normalized anisotropy fields are $%
h_{k1}=2K_{1}/M_{s}^{2}$ and $h_{k2}=2K_{2}/M_{s}^{2}$. The normalized
exchange field, assuming an interaction strength $A$ (acting uniformly
between grain centers)$,$ is $h_{ex}=2A/D^{2}M_{s}^{2}$ (although here
normalization is by $M_{s}^{2}$ rather than by $K$ typical in micromagnetic
simulations of recording media\cite{bertzhu}). Normalization by $M_{s}^{2}V$
yields the magnetostatic term (using a dipole approximation) without any
multiplicative constant. The magnetostatic and exchange interactions are
assumed to be small compared to the anisotropy energies $h_{k1},h_{k2}\gg
1\gg h_{ex}.$ With this assumption the equilibrium state is (Fig.1):

\begin{equation}
\overrightarrow{m}_{1}=e_{z},\ \ \overrightarrow{m}_{2}=e_{z}  \label{equil}
\end{equation}
For small excitations around equilibrium ($e_{z}$), we only need consider
second order variations so that:

\begin{eqnarray}
m_{1z} &=&1-\frac{1}{2}(m_{1x}^{2}+m_{1y}^{2})  \label{second} \\
m_{1z}m_{2z} &=&1-\frac{1}{2}(m_{1x}^{2}+m_{1y}^{2}+m_{2x}^{2}+m_{2y}^{2}) 
\nonumber
\end{eqnarray}
The magnetic energy of the host grains, to second order, is:

\begin{eqnarray}
E_{m} &=&M_{s}^{2}V[-\frac{1}{2}(-h_{k1}+1-h_{ex})(m_{1x}^{2}+m_{1y}^{2})-%
\frac{1}{2}(-h_{k2}+1-h_{ex})(m_{2x}^{2}+m_{2y}^{2})  \label{enequa} \\
&&+(1-h_{ex})m_{1y}m_{2y}-(2+h_{ex})m_{1x}m_{2x}]  \nonumber
\end{eqnarray}
For each grain ($j$), we transform the two magnetization components
orthogonal to the equilibrium direction ( $m_{jx}^{{}},m_{jy}^{{}})$ into
(linearized) rotating magnetization components ($a_{j}^{\ast },a_{j})$ (e.g. 
\cite{spark}):

\begin{equation}
a_{j}=\frac{m_{jx}+im_{jy}^{{}}}{\sqrt{2}},\text{ }a_{j}^{\ast }=\frac{%
m_{jx}^{{}}-im_{jy}^{{}}}{\sqrt{2}}  \label{hols}
\end{equation}
In these coordinates, again in the lowest order quadratic variation, the
magnetic energy (\ref{enequa}) can be written as: \ 
\begin{eqnarray}
E_{m} &=&M_{s}^{2}V[-(-h_{k1}+1-h_{ex})a_{1}a_{1}^{\ast
}-(-h_{k2}+1-h_{ex})a_{2}a_{2}^{\ast }  \label{cantran} \\
&&+1/2(1-h_{ex})(a_{1}a_{2}^{\ast }+a_{1}^{\ast
}a_{2}-a_{1}a_{2}-a_{1}^{\ast }a_{2}^{\ast })  \nonumber \\
&&-1/2(2+h_{ex})(a_{1}a_{2}+a_{1}^{\ast }a_{2}^{\ast }+a_{1}a_{2}^{\ast
}+a_{1}^{\ast }a_{2}^{\ast })]  \nonumber
\end{eqnarray}

In order to consider localized relaxation processes the host magnetic grains
are bilinearly coupled to two separate oscillating thermal baths\cite{scully}%
. This is a simplified model for local relaxation processes of the
interacting magnetization system. A \ physical example would be relaxation
by localized high moment Rare Earth impurities (e.g. see\cite{ReNiFe}). The
total energy including magnetic energy, thermal bath energy and interaction
energy is:

\begin{eqnarray}
E_{t} &=&E_{m}+E_{b}+E_{I}  \label{sysene} \\
E_{m} &=&E_{m}(a_{1,}a_{1}^{\ast },a_{2},a_{2}^{\ast })  \nonumber \\
E_{b} &=&\frac{M_{s}V}{\gamma }%
\mathrel{\mathop{\sum }\limits_{k}}%
\omega _{1k}b_{1k}b_{1k}^{\ast }+\frac{M_{s}V}{\gamma }%
\mathrel{\mathop{\sum }\limits_{k}}%
\omega _{2k}b_{2k}b_{2k}^{\ast }  \nonumber \\
E_{I} &=&\frac{M_{s}V}{\gamma }%
\mathrel{\mathop{\sum }\limits_{k}}%
g_{1k}(b_{1k}a_{1}^{\ast }+b_{1k}^{\ast }a_{1})+\frac{M_{s}V}{\gamma }%
\mathrel{\mathop{\sum }\limits_{k}}%
g_{2k}(b_{2k}a_{2}^{\ast }+b_{2k}^{\ast }a_{2})  \nonumber
\end{eqnarray}
where $b_{jk},b_{jk}^{\ast }$ are the oscillating thermal bath variables. $%
g_{1k},g_{2k}$ represent the coupling strength of the magnetic systems and
the thermal baths. Note that the bath terms are in the form of independent
harmonic oscillators. Using the transformation (\ref{hols}), the interaction
term $E_{I}$ can be viewed as simply a Zeeman coupling with a thermal field.
Here following \cite{scully}, we assume that thermal bath is in equilibrium
and the magnetic coupling is only a small perturbation to thermal baths.

\section{Model Dynamic Equations}

The standard procedure is to obtain a closed dynamic equation in the
magnetization variables by explicitly eliminating the thermal bath variables 
\cite{scully}. The Hamiltonian equations for the magnetic system are:

\begin{eqnarray}
\frac{da_{j}}{dt} &=&-i\frac{\gamma }{M_{s}V}\frac{\partial E}{\partial
a_{j}^{\ast }}  \label{dyna} \\
\frac{da_{j}^{\ast }}{dt} &=&i\frac{\gamma }{M_{s}V}\frac{\partial E}{%
\partial a_{j}}  \nonumber
\end{eqnarray}
which can be written in matrix form as:

\begin{equation}
\frac{d}{dt}\left( 
\begin{array}{c}
a_{1} \\ 
a_{1}^{\ast } \\ 
a_{2} \\ 
a_{2}^{\ast }
\end{array}
\right) =iG\left( 
\begin{array}{c}
a_{1} \\ 
a_{1}^{\ast } \\ 
a_{2} \\ 
a_{2}^{\ast }
\end{array}
\right) +i\left( 
\begin{array}{c}
\mathrel{\mathop{-\sum }\limits_{k}}%
g_{1k}b_{1k} \\ 
\mathrel{\mathop{\sum }\limits_{k}}%
g_{1k}b_{1k}^{\ast } \\ 
\mathrel{\mathop{-\sum }\limits_{k}}%
g_{2k}b_{2k} \\ 
\mathrel{\mathop{\sum }\limits_{k}}%
g_{2k}b_{2k}^{\ast }
\end{array}
\right)  \label{msd}
\end{equation}
where:

\begin{equation}
\text{\ }G=\left( 
\begin{array}{cccc}
(-h_{k1}+1-h_{ex}) & 0 & (1+2h_{ex})/2 & 3/2 \\ 
0 & -(-h_{k1}+1-h_{ex}) & -3/2 & -(1+2\ast h_{ex})/2 \\ 
(1+2h_{ex})/2 & 3/2 & (-h_{k2}+1-h_{ex}) & 0 \\ 
-3/2 & -(1+2\ast h_{ex})/2 & 0 & -(-h_{k2}+1-h_{ex})
\end{array}
\right) \text{\ }\gamma M_{s}\text{\ \ }  \label{gymat}
\end{equation}
is the matrix for gyromagnetic precession.

In order to obtain a closed equation for magnetic system variables $%
(a_{1},a_{1}^{\ast },a_{2},a_{2}^{\ast })$, we need to formally represent $%
(b_{1k},b_{1k}^{\ast },b_{2k},b_{2k}^{\ast })$ in terms of $%
(a_{1},a_{1}^{\ast },a_{2},a_{2}^{\ast })$ in (\ref{msd}). We expect the
closed equation for $(a_{1},a_{1}^{\ast },a_{2},a_{2}^{\ast })$ to be a
stochastic differential equation under a Markov approximation. However,
before starting to represent $(b_{1k},b_{1k}^{\ast },b_{2k},b_{2k}^{\ast })$
in terms of $(a_{1},a_{1}^{\ast },a_{2},a_{2}^{\ast })$, we notice that (\ref
{msd}) has mixed times because of nondiagonal terms in the matrix (\ref
{gymat}). Thus, due to intergranular interactions, the distinct magnetic
system time scales are not well presented in the rotating magnetization
components $(a_{1},a_{1}^{\ast },a_{2},a_{2}^{\ast }).$ In the Markov
approximation for stochastic dynamics, distinct system time scales must be
separated from thermal bath time scales. So we first need to obtain
explicitly system time scales \cite{Breuer}. This is done by normal mode
analysis (e.g.\cite{spark}).

The nondiagonal matrix (\ref{gymat}) can be diagonalized into the following
form:

\begin{eqnarray}
\text{\ }G &=&U\Lambda U^{-1}\text{\ \ \ }  \label{diag} \\
U &=&\left( 
\begin{array}{cccc}
u_{11} & v_{11} & u_{12} & v_{12} \\ 
v_{11} & u_{11} & v_{12} & u_{12} \\ 
u_{21} & v_{21} & u_{22} & v_{22} \\ 
v_{21} & u_{21} & v_{22} & u_{22}
\end{array}
\right) ,\Lambda =\left( 
\begin{array}{cccc}
-\omega _{1} & 0 & 0 & 0 \\ 
0 & \omega _{1} & 0 & 0 \\ 
0 & 0 & -\omega _{2} & 0 \\ 
0 & 0 & 0 & \omega _{2}
\end{array}
\right)  \nonumber
\end{eqnarray}
Thus, the system gyromagnetic motion alone without a thermal bath or
equivalently a relaxation mechanism can be written as:

\begin{eqnarray}
\frac{d}{dt}\left( 
\begin{array}{c}
c_{1} \\ 
c_{1}^{\ast } \\ 
c_{2} \\ 
c_{2}^{\ast }
\end{array}
\right) &=&\left( 
\begin{array}{cccc}
-i\omega _{1} & 0 & 0 & 0 \\ 
0 & i\omega _{1} & 0 & 0 \\ 
0 & 0 & -i\omega _{2} & 0 \\ 
0 & 0 & 0 & i\omega _{2}
\end{array}
\right) \left( 
\begin{array}{c}
c_{1} \\ 
c_{1}^{\ast } \\ 
c_{2} \\ 
c_{2}^{\ast }
\end{array}
\right)  \label{normtion} \\
\left( 
\begin{array}{c}
a_{1} \\ 
a_{1}^{\ast } \\ 
a_{2} \\ 
a_{2}^{\ast }
\end{array}
\right) &=&\left( 
\begin{array}{cccc}
u_{11} & v_{11} & u_{12} & v_{12} \\ 
v_{11} & u_{11} & v_{12} & u_{12} \\ 
u_{21} & v_{21} & u_{22} & v_{22} \\ 
v_{21} & u_{21} & v_{22} & u_{22}
\end{array}
\right) \left( 
\begin{array}{c}
c_{1} \\ 
c_{1}^{\ast } \\ 
c_{2} \\ 
c_{2}^{\ast }
\end{array}
\right)  \nonumber
\end{eqnarray}
1where $(c_{1},c_{1}^{\ast },c_{2},c_{2}^{\ast })$ are the system normal
modes and distinct system times scales can be determined as $1/\omega _{1}$%
and $1/\omega _{2}.$ $\omega _{1}$and $\omega _{2}$ are the magnetic system
resonant frequencies.

Now we represent total Hamiltonian (\ref{enequa}) using normal mode
coordinate:

\begin{eqnarray}
E &=&\frac{M_{s}V}{\gamma }[\omega _{1}c_{1}c_{1}^{\ast }+\omega
_{2}c_{2}c_{2}^{\ast }]+\frac{M_{s}V}{\gamma }%
\mathrel{\mathop{\sum }\limits_{k}}%
\omega _{1k}b_{1k}b_{1k}^{\ast }+\frac{M_{s}V}{\gamma }%
\mathrel{\mathop{\sum }\limits_{k}}%
\omega _{2k}b_{2k}b_{2k}^{\ast }  \label{nornene} \\
&&+\frac{M_{s}V}{\gamma }\sum \left( 
\begin{array}{c}
c_{1}(g_{1k}v_{11}b_{1k}+g_{1k}u_{11}b_{1k}^{\ast
}+g_{2k}v_{21}b_{2k}+g_{2k}u_{21}b_{2k}^{\ast }) \\ 
c_{1}^{\ast }(g_{1k}u_{11}b_{1k}+g_{1k}v_{11}b_{1k}^{\ast
}+g_{2k}u_{21}b_{2k}+g_{2k}v_{21}b_{2k}^{\ast }) \\ 
c_{2}(g_{1k}v_{12}b_{1k}+g_{1k}u_{12}b_{1k}^{\ast
}+g_{2k}v_{22}b_{2k}+g_{2k}u_{22}b_{2k}^{\ast }) \\ 
c_{2}^{\ast }(g_{1k}u_{12}b_{1k}+g_{1k}v_{12}b_{1k}^{\ast
}+g_{2k}u_{22}b_{2k}+g_{2k}v_{22}b_{2k}^{\ast })
\end{array}
\right)  \nonumber
\end{eqnarray}
It should be pointed out here that diagonalizing the matrix $G$ in (\ref
{diag}) together with the requirement that the normal mode energy in form of 
$E=\frac{M_{s}V}{\gamma }[\omega _{1}c_{1}c_{1}^{\ast }+\omega
_{2}c_{2}c_{2}^{\ast }]$ (\ref{nornene}) determines the transformation
matrix $U$ uniquely. The dynamic equations for the magnetic system can be
obtained from the Hamiltonian:

\begin{equation}
\frac{dc_{i}}{dt}=-i\frac{\gamma }{M_{s}V}\frac{\partial E}{\partial
c_{i}^{\ast }}  \label{cnr}
\end{equation}
so that:

\begin{eqnarray}
\frac{dc_{1}}{dt} &=&-i\omega _{1}c_{1}-i%
\mathrel{\mathop{\sum }\limits_{k}}%
(g_{1k}u_{11}b_{1k}+g_{1k}v_{11}b_{1k}^{\ast
}+g_{2k}u_{21}b_{2k}+g_{2k}v_{21}b_{2k}^{\ast })  \label{eqc} \\
\frac{dc_{2}}{dt} &=&-i\omega _{2}c_{2}-i%
\mathrel{\mathop{\sum }\limits_{k}}%
(g_{1k}u_{12}b_{1k}+g_{1k}v_{12}b_{1k}^{\ast
}+g_{2k}u_{22}b_{2k}+g_{2k}v_{22}b_{2k}^{\ast })  \nonumber
\end{eqnarray}
The dynamic equations for bath variables are:

\begin{equation}
\frac{db_{ik}}{dt}=-i\frac{\gamma }{M_{s}V}\frac{\partial E}{\partial
b_{ik}^{\ast }}  \label{bnr}
\end{equation}
that is:

\begin{eqnarray}
\frac{db_{1k}}{dt} &=&-i\omega
_{1k}b_{1k}-i(g_{1k}u_{11}c_{1}+g_{1k}v_{11}c_{1}^{\ast
}+g_{1k}u_{12}c_{2}+g_{1k}v_{12}c_{2}^{\ast })  \label{eqb} \\
\frac{db_{2k}}{dt} &=&-i\omega
_{2}b_{2k}-i(g_{2k}u_{21}c_{1}+g_{2k}v_{21}c_{1}^{\ast
}+g_{2k}u_{22}c_{2}+g_{2k}v_{22}c_{2}^{\ast })  \nonumber
\end{eqnarray}
Equation (\ref{eqb}) can be formally solved as: 
\begin{eqnarray}
b_{1k}(t) &=&b_{1k}(0)e^{-i\omega _{1k}t}-ig_{1k}\int_{0}^{t}dt^{\prime
}(u_{11}c_{1}+v_{11}c_{1}^{\ast }+u_{12}c_{2}+v_{12}c_{2}^{\ast
})e^{-i\omega _{1k}(t-t^{\prime })}  \label{solb} \\
b_{2k}(t) &=&b_{2k}(0)e^{-i\omega _{2k}t}-ig_{2k}\int_{0}^{t}dt^{\prime
}(u_{21}c_{1}+v_{21}c_{1}^{\ast }+u_{22}c_{2}+v_{22}c_{2}^{\ast
})e^{-i\omega _{2k}(t-t^{\prime })}  \nonumber
\end{eqnarray}

In order to simplify the calculation, in the following we assume the two
thermal baths are identical except for a weighting factor in coupling
strength to the magnetic system:

\begin{equation}
g_{2k}=\beta g_{1k},\text{ }\omega _{1k}=\omega _{2k}  \label{wei}
\end{equation}
Substituting (\ref{solb}) into (\ref{eqc}), we obtain:

\begin{eqnarray}
\frac{dc_{1}}{dt} &=&-i\omega _{1}c_{1}-i%
\mathrel{\mathop{\sum }\limits_{k}}%
g_{1k}[u_{11}b_{1k}(0)+\beta u_{21}b_{2k}(0)]e^{-i\omega _{k}t}
\label{clong} \\
&&-i%
\mathrel{\mathop{\sum }\limits_{k}}%
g_{1k}[v_{11}b_{1k}^{\ast }(0)+\beta v_{21}b_{2k}^{\ast }(0)]e^{i\omega
_{k}t}  \nonumber \\
&&-%
\mathrel{\mathop{\sum }\limits_{k}}%
g_{1k}^{2}\int_{0}^{t}dt^{\prime }e^{-i\omega _{k}(t-t^{\prime
})}(u_{11}^{2}c_{1}+u_{11}v_{11}c_{1}^{\ast
}+u_{11}u_{12}c_{2}+u_{11}v_{12}c_{2}^{\ast })  \nonumber \\
&&-%
\mathrel{\mathop{\sum }\limits_{k}}%
\beta ^{2}g_{1k}^{2}\int_{0}^{t}dt^{\prime }e^{-i\omega _{k}(t-t^{\prime
})}(u_{21}^{2}c_{1}+u_{21}v_{21}c_{1}^{\ast
}+u_{21}u_{22}c_{2}+u_{21}v_{22}c_{2}^{\ast })  \nonumber \\
&&+%
\mathrel{\mathop{\sum }\limits_{k}}%
g_{1k}^{2}\int_{0}^{t}dt^{\prime }e^{i\omega _{k}(t-t^{\prime
})}(v_{11}^{2}c_{1}+v_{11}u_{11}c_{1}^{\ast
}+v_{11}v_{12}c_{2}+v_{11}u_{12}c_{2}^{\ast })  \nonumber \\
&&+%
\mathrel{\mathop{\sum }\limits_{k}}%
\beta ^{2}g_{1k}^{2}\int_{0}^{t}dt^{\prime }e^{-i\omega _{k}(t-t^{\prime
})}(v_{21}^{2}c_{1}+v_{21}u_{21}c_{1}^{\ast
}+v_{21}v_{22}c_{2}+v_{21}u_{22}c_{2}^{\ast })  \nonumber
\end{eqnarray}
and similarly for $c_{2}.$

\section{\protect\bigskip Stochastic Dynamic Equations}

We assume that the thermal bath is in thermal equilibrium with a continuum
density of states: 
\begin{eqnarray}
&<&b_{ik}(0)>=<b_{ik}^{\ast }(0)>=0  \label{ebe} \\
&<&b_{ik}(0)b_{ik^{\prime }}^{\ast }(0)>=\delta _{kk^{\prime }}n_{k} 
\nonumber
\end{eqnarray}
where $i=1,2$, $n_{k}$ is the energy of the $kth$ component of the
oscillating thermal bath, proportional to temperature if the thermal bath is
in equilibrium. The summation over discrete energy levels can be
approximated by a continuous integration:

\begin{equation}
\mathrel{\mathop{\sum }\limits_{k}}%
g_{1k}^{2}\rightarrow \int d\omega _{k}D(\omega _{k})g^{2}(\omega _{k})
\label{dens}
\end{equation}
and where occupation number in (\ref{ebe}) is directly proportional to the
temperature: $n_{k}\thicksim k_{B}T$.

We need to single out the magnetic system time scales from the coupled total
system time scales. This is done by the following transformation\cite{scully}%
:

\begin{equation}
\widetilde{c}_{1}(t)=c_{1}(t)e^{i\omega _{1}t},\ \widetilde{c}%
_{2}(t)=c_{2}(t)e^{i\omega _{2}t}  \label{timelong}
\end{equation}
Substituting (\ref{dens}) and (\ref{timelong}) into (\ref{clong}),\ we
obtain:

\begin{equation}
\begin{array}{c}
\frac{d\widetilde{c}_{1}}{dt}=f_{1}(t)-\int d\omega _{k}D(\omega
_{k})g^{2}(\omega _{k})\int_{0}^{t}dt^{\prime } \\ 
\left[ 
\begin{array}{c}
(u_{11}^{2}+\beta ^{2}u_{21}^{2})\widetilde{c}_{1}(t^{\prime })e^{-i(\omega
_{k}-\omega _{1})(t-t^{\prime })}+(u_{11}v_{11}+\beta ^{2}u_{21}v_{21})%
\widetilde{c}_{1}^{\ast }(t^{\prime })e^{-i(\omega _{k}+\omega
_{1})(t-t^{\prime })}e^{2i\omega _{1}t} \\ 
(u_{11}u_{12}+\beta ^{2}u_{21}u_{22})\widetilde{c}_{2}(t^{\prime
})e^{-i(\omega _{k}-\omega _{1})(t-t^{\prime })}e^{i(\omega _{1}-\omega
_{2})t}+(u_{11}v_{12}+\beta ^{2}u_{21}v_{22})\widetilde{c}_{2}(t^{\prime
})e^{-i(\omega _{k}+\omega _{1})(t-t^{\prime })}e^{i(\omega _{1}+\omega
_{2})t} \\ 
-(v_{11}^{2}+\beta ^{2}v_{21}^{2})\widetilde{c}_{1}(t^{\prime })e^{i(\omega
_{k}+\omega _{1})(t-t^{\prime })}-(v_{11}u_{11}+\beta ^{2}v_{21}v_{22})%
\widetilde{c}_{1}^{\ast }(t^{\prime })e^{i(\omega _{k}-\omega
_{1})(t-t^{\prime })}e^{2i\omega _{1}t} \\ 
-(v_{11}v_{12}+\beta ^{2}v_{21}v_{22})\widetilde{c}_{2}(t^{\prime
})e^{i(\omega _{k}+\omega _{2})(t-t^{\prime })}e^{i(\omega _{1}-\omega
_{2})t}-(v_{11}u_{12}+\beta ^{2}v_{21}u_{22})\widetilde{c}_{2}^{\ast
}(t^{\prime })e^{i(\omega _{k}+\omega _{2})(t-t^{\prime })}e^{i(\omega
_{1}+\omega _{2})t}
\end{array}
\right]
\end{array}
\label{fiave}
\end{equation}
where

\begin{eqnarray}
f_{1}(t) &=&-i%
\mathrel{\mathop{\sum }\limits_{k}}%
g_{1k}[u_{11}b_{1k}(0)+\beta u_{21}b_{2k}(0)]e^{-i(\omega _{k}-\omega _{1})t}
\label{noise} \\
&&-i%
\mathrel{\mathop{\sum }\limits_{k}}%
g_{1k}[v_{11}b_{1k}^{\ast }(0)+\beta v_{21}b_{2k}^{\ast }(0)]e^{i(\omega
_{k}+\omega _{1})t}  \nonumber
\end{eqnarray}
is the thermal fluctuation field for the first mode. It satisfies:

\begin{eqnarray}
&<&f_{1}(t)>=0  \label{thermal fluc} \\
&<&f_{1}(t)f_{1}^{\ast }(t^{\prime })>=D(\omega _{1})g^{2}(\omega
_{1})n(\omega _{1})(u_{11}^{2}+\beta ^{2}u_{21}^{2})\delta (t-t^{\prime }) 
\nonumber
\end{eqnarray}
A similar expression for the second mode can be obtained.

In the Markovian approximation \cite{scully},\cite{Breuer}, we neglect
memory for the slow variable ($\widetilde{c}_{1}(t^{\prime })\rightarrow $ $%
\widetilde{c}_{1}(t)$). For long times $t>>1/(\omega _{1}-\omega _{k})$ the
upper integral limit may be put to infinity and using:

\begin{equation}
\int_{0}^{\infty }dxe^{i\omega x}=\pi \delta (\omega )+iP.V.(\frac{1}{\omega 
})  \label{pv}
\end{equation}
only the following terms remain:

\begin{equation}
\begin{array}{c}
\frac{d\widetilde{c}_{1}}{dt}=f_{1}(t)e^{i\omega _{1}t}-\int d\omega
_{k}D(\omega _{k})g^{2}(\omega _{k})\int_{0}^{\infty }dt^{\prime } \\ 
\left[ 
\begin{array}{c}
(u_{11}^{2}+\beta ^{2}u_{21}^{2})\widetilde{c}_{1}(t)e^{-i(\omega
_{k}-\omega _{1})(t-t^{\prime })} \\ 
(u_{11}u_{12}+\beta ^{2}u_{21}u_{22})\widetilde{c}_{2}(t)e^{-i(\omega
_{k}-\omega _{1})(t-t^{\prime })}e^{i(\omega _{1}-\omega _{2})t} \\ 
-(v_{11}u_{11}+\beta ^{2}v_{21}v_{22})\widetilde{c}_{1}^{\ast
}(t)e^{i(\omega _{k}-\omega _{1})(t-t^{\prime })}e^{2i\omega _{1}t}
\end{array}
\right]
\end{array}
\label{remain}
\end{equation}
Because $e^{2i\omega _{1}t}$ is a fast oscillating term\ for $t>>1/\omega
_{1}$, the $-(v_{11}u_{11}+\beta ^{2}v_{21}v_{22})\widetilde{c}_{1}^{\ast
}(t)e^{i(\omega _{k}-\omega _{1})(t-t^{\prime })}e^{2i\omega _{1}t}$ term
can be included into the fluctuating term. This is the rotational wave
approximation \cite{Breuer}. If the gyromagnetic system is not degenerate ($%
\omega _{1}\neq $ $\omega _{2}$) and the gyromagnetic rotating frequency gap
is sufficiently large $t>>1/(\omega _{1}-\omega _{2})$, $e^{i(\omega
_{1}-\omega _{2})t}$ is also a fast oscillating term and the $%
(u_{11}u_{12}+\beta ^{2}u_{21}u_{22})\widetilde{c}_{2}(t)e^{-i(\omega
_{k}-\omega _{1})(t-t^{\prime })}e^{i(\omega _{1}-\omega _{2})t}$ term also
enters into the fluctuating terms. \ As discussed in section V, for a wide
range of parameters, even for identical grains, nondegenerate gyromagnetic
rotation is guaranteed.

The following stochastic differential equation for $\widetilde{c}_{1}(t)$ is
obtained:

\begin{equation}
\frac{d\widetilde{c}_{1}}{dt}=\widetilde{F}_{1}(t)-D(\omega
_{1})g^{2}(\omega _{1})(u_{11}^{2}+\beta ^{2}u_{21}^{2})\widetilde{c}_{1}(t)
\label{cti}
\end{equation}
where the damping term is defined as:

\begin{equation}
\eta _{1}=D(\omega _{1})g^{2}(\omega _{1})(u_{11}^{2}+\beta ^{2}u_{21}^{2})
\label{F1DAMP}
\end{equation}
Weak interactions between the system and the thermal reservoir is usually
assumed for a Markovian approximation and this corresponds to $\eta _{1}<<1$%
. The thermal fluctuation term(\ref{cti}) is:

\begin{eqnarray}
\widetilde{F}_{1}(t) &=&-i%
\mathrel{\mathop{\sum }\limits_{k}}%
g_{1k}\{[u_{11}b_{1k}(0)+\beta u_{21}b_{2k}(0)]e^{-i(\omega _{k}-\omega
_{1})t}  \label{F1noise} \\
&&+[v_{11}b_{1k}^{\ast }(0)+\beta v_{21}b_{2k}^{\ast }(0)]e^{i(\omega
_{k}+\omega _{1})t}\}  \nonumber \\
&&-\int d\omega _{k}D(\omega _{k})g^{2}(\omega _{k})\int_{0}^{\infty
}dt(u_{11}u_{12}+\beta ^{2}u_{21}u_{22})\widetilde{c}_{2}(t)e^{-i(\omega
_{k}-\omega _{1})(t-t^{\prime })}e^{i(\omega _{1}-\omega _{2})t}  \nonumber
\\
&&+\int d\omega _{k}D(\omega _{k})g^{2}(\omega _{k})\int_{0}^{\infty
}dt^{\prime }(v_{11}u_{11}+\beta ^{2}v_{21}v_{22})\widetilde{c}_{1}^{\ast
}(t)e^{i(\omega _{k}-\omega _{1})(t-t^{\prime })}e^{2i\omega _{1}t} 
\nonumber
\end{eqnarray}
Notice that the last two terms are from the fast oscillating terms and have
the magnitude proportional to $\eta _{1}.$ The thermal fluctuation has zero
mean and, to leading order (neglecting $\eta _{1}^{2}$), the variance is:

\begin{mathletters}
\begin{equation}
<\widetilde{F}_{1}(t)\widetilde{F}_{1}^{\ast }(t^{\prime })>=D(\omega
_{1})g^{2}(\omega _{1})n(\omega _{1})(u_{11}^{2}+\beta ^{2}u_{21}^{2})\delta
(t-t^{\prime })  \label{var}
\end{equation}
A similar equation for $\widetilde{c}_{2}(t)$ can be obtained based upon the
same conditions:

\end{mathletters}
\begin{eqnarray}
\frac{d\widetilde{c}_{2}}{dt} &=&\widetilde{F}_{2}(t)-\eta _{2}\widetilde{c}%
_{1}(t)  \label{C2} \\
\eta _{2} &=&D(\omega _{2})g^{2}(\omega _{2})(\beta
^{2}u_{22}^{2}+u_{12}^{2})  \nonumber \\
\widetilde{F}_{2}(t) &=&-i%
\mathrel{\mathop{\sum }\limits_{k}}%
g_{1k}\{[u_{12}b_{1k}(0)+\beta u_{22}b_{2k}(0)]e^{-i(\omega _{k}-\omega
_{2})t}  \nonumber \\
&&+[v_{12}b_{1k}^{\ast }(0)+\beta v_{22}b_{2k}^{\ast }(0)]e^{i(\omega
_{k}+\omega _{2})t}\}  \nonumber \\
&&+higher\text{ }order\text{ }fast\text{ }oscillating\text{ }terms  \nonumber
\end{eqnarray}
To summarize, the thermal fluctuation in the two modes have to leading order
the following properties:

\begin{eqnarray}
&<&\widetilde{F}_{1}(t)>=0,<\widetilde{F}_{2}(t)>=0  \label{NOITERM} \\
&<&\widetilde{F}_{1}(t)\widetilde{F}_{1}^{\ast }(t^{\prime })>=D(\omega
_{1})g^{2}(\omega _{1})n(\omega _{1})(u_{11}^{2}+\beta ^{2}u_{21}^{2})\delta
(t-t^{\prime })  \nonumber \\
&<&\widetilde{F}_{2}(t)\widetilde{F}_{2}^{\ast }(t^{\prime })>=D(\omega
_{2})g^{2}(\omega _{2})n(\omega _{2})(\beta ^{2}u_{22}^{2}+u_{12}^{2})\delta
(t-t^{\prime })  \nonumber \\
&<&\widetilde{F}_{1}(t)\widetilde{F}_{2}^{\ast }(t^{\prime })>=0  \nonumber
\end{eqnarray}

The last condition in (\ref{NOITERM}) gives uncorrelated thermal
fluctuations in the two normal modes. It should be pointed out that in
principle the fast oscillating terms in the thermal noise could result in
correlation between $\widetilde{F}_{1}(t)$ and $\widetilde{F}_{2}(t)$.
However, those terms are of order $\eta ^{2}$ and can be neglected for a
consistent Markovian approximation with weak interactions between the
magnetic system and thermal reservoir.

Transforming (\ref{cti}) and (\ref{C2}) into $c_{1}(t)$ and $c_{2}(t)$
coordinates, we obtain the following stochastic equations for the collective
normal modes of the interacting magnetic system:

\begin{eqnarray}
\frac{dc_{1}}{dt}+\eta _{1}c_{1} &=&-i\omega _{1}c_{1}+f_{1}(t)  \label{fff}
\\
\frac{dc_{2}}{dt}+\eta _{2}c_{2} &=&-i\omega _{2}c_{2}+f_{2}(t)  \nonumber \\
\eta _{1} &=&D(\omega _{1})g^{2}(\omega _{1})(u_{11}^{2}+\beta
^{2}u_{21}^{2})  \nonumber \\
\eta _{2} &=&D(\omega _{2})g^{2}(\omega _{2})(\beta
^{2}u_{22}^{2}+u_{12}^{2})  \nonumber \\
&<&f_{1}(t)>=0,<f_{2}(t)>=0  \nonumber \\
&<&f_{1}(t)f_{1}^{\ast }(t^{\prime })>=D(\omega _{1})g^{2}(\omega
_{1})n(\omega _{1})(u_{11}^{2}+\beta ^{2}u_{21}^{2})\delta (t-t^{\prime }) 
\nonumber \\
&<&f_{2}(t)f_{2}^{\ast }(t^{\prime })>=D(\omega _{2})g^{2}(\omega
_{2})n(\omega _{2})(\beta ^{2}u_{22}^{2}+u_{12}^{2})\delta (t-t^{\prime }) 
\nonumber \\
&<&f_{1}(t)f_{2}^{\ast }(t^{\prime })>=0  \nonumber
\end{eqnarray}

Thus, this analysis without any a {\it priori} assumptions has given
stochastic differential equations for the two independent normal modes in
the form of independent damped harmonic oscillators driven by uncorrelated
thermal fluctuations. Note that the damping terms $\eta _{1}\left( \omega
_{1}\right) $ and $\eta _{2}\left( \omega _{2}\right) $ are directly
proportional to thermal variance terms $<f_{1}(t)f_{1}^{\ast }(t^{\prime
})>=D_{c1}$ and $<f_{2}(t)f_{2}^{\ast }(t^{\prime })>=D_{c2}$ respectively.
The ratio of fluctuation magnitudes in two modes are: 
\begin{equation}
\frac{D_{c1}}{D_{c2}}=\frac{D(\omega _{1})g^{2}(\omega _{1})n(\omega
_{1})(u_{11}^{2}+\beta ^{2}u_{21}^{2})}{D(\omega _{2})g^{2}(\omega
_{2})n(\omega _{2})(\beta ^{2}u_{22}^{2}+u_{12}^{2})}  \label{ratio2}
\end{equation}
For a magnetic system that satisfies a canonical distribution around
equilibrium, the magnitude of the damping coefficients $\eta _{1}$ and $\eta
_{2}$ are related to the thermal fluctuations terms $D_{c1}$ and $D_{c2}$
through the fluctuation-dissipation condition \cite{scully}:

\begin{eqnarray}
\eta _{1} &=&\frac{D_{c1}M_{s}V}{\gamma k_{B}T}\omega _{1}=\alpha
_{Q1}\omega _{1}  \label{gerd} \\
\eta _{2} &=&\frac{D_{c2}M_{s}V}{\gamma k_{B}T}\omega _{2}=\alpha
_{Q2}\omega _{2}  \nonumber
\end{eqnarray}
where $\eta _{1}$ and $\eta _{2}$ are damping rate.

\section{Discussion}

We have shown here (\ref{fff}) that local physical relaxation mechanisms
give collective stochastic dynamics in the system normal modes for
interacting magnetic grains or continuum discretization cells. For a given
system the specific damping terms can be evaluated using (\ref{gerd}). These
results can be generalized for any system of $N$ interacting magnetic
sub-units. The general operative approach is to first diagonalize the ($%
2N\times 2N$) matrix of the gyromagnetic precession (near equilibrium)
without any damping terms to find the collective modes and their eigenvalues
or resonance frequencies. Following that, using expanded forms of (\ref{gerd}%
), damping and thermal fields are added to each of the collective mode
dynamics. The total energy in each mode is $k_{B}T$ and the
fluctuation-dissipation theorem relates the variance of the thermal term to
the damping. As will be shown in the specific examples below, there is no
contradiction between local relaxation mechanisms and collective stochastic
dynamics. The interactions between the elements allow damping in one cell to
be felt by all the others. The interactions also cause the dynamics to
reflect the overall sample geometry or anisotropy.

Here we evaluate some examples for this two-grain system. The most general
case of the dependence of\ the collective stochastic dynamics on interaction
strength can be obtained from (\ref{diag}). We \ begin with the results for
a system of identical grains with no exchange $\left( h_{ex}=0\right) .$ The
mode resonance frequencies are: 
\begin{eqnarray}
\frac{\omega _{1}}{\gamma H_{K1}} &=&\sqrt{1-3\frac{M_{s}}{H_{K1}}}
\label{resf} \\
\frac{\omega _{2}}{\gamma H_{K1}} &=&\sqrt{1-\frac{M_{s}}{H_{K1}}-2\left( 
\frac{M_{s}}{H_{K1}}\right) ^{2}}  \nonumber
\end{eqnarray}
\qquad\ In Fig.2a these normalized frequencies are potted versus interaction
strength\ $M_{s}/H_{K1}$. With weak or vanishing magnetostatic interactions
the frequencies are simply that of independent grains with $\omega
_{1}=\gamma H_{K1}$ and $\omega _{2}=\gamma H_{K2}=\gamma H_{k1}.$ With
increasing interaction both resonance frequencies decrease. The modes in
zero field are asymmetric coherent rotation (mode 1) and asymmetric fanning
(mode 2)$\cite{xyz}.$ The lowest frequency mode is almost coherent, because
there is less magnetostatic energy to rotate into the particle axis
direction ($x$: Fig.1). As the magnetostatic interaction is increased the
energy barrier (or quadratic curvature) decreases and at $%
M_{s}/H_{K1}\approx 0.33$ the frequency vanishes and the equilibrium
direction moves from the ``$z"$ direction to be along the line joining the
particles (the ``$x$''\ direction). For the higher energy mode the frequency 
$\omega _{2}$ also decreases with interaction strength and eventually the
energy barrier will vanish $\left( \omega _{2}\rightarrow 0\right) $, but at
a higher $M_{s}/H_{K1}=0.5$ due to the larger magnetostatic energy of that
mode.

In Fig.2b normalized resonance frequencies are plotted versus exchange $%
h_{ex}$ for the case $M_{s}/H_{K1}=0.09$ and $H_{K2}=1.95H_{K1}.$ At $%
h_{ex}=0$ the resonance frequencies are almost the ratio of the $H_{K}$
values. The quasi-coherent mode frequency $(\omega _{1})$ hardly varies with
exchange, as expected, since coherent motion does not involve exchange
energy. The incoherent mode frequency $\omega _{2}$ increases rapidly with
exchange because of the increased exchange energy of that non-uniform mode.
We emphasize that the stochastic dynamic modeling in this paper is for $%
\omega _{2}\neq \omega _{1}$(and sufficiently different). However, the plots
shown here indicate that for almost all cases of interest, this condition
holds, even for identical particles with finite coupling.

We now explore the variation of the mode damping parameters with
magnetostatic interaction for $h_{ex}=0$, as in Fig.2a and with slightly
differing anisotropy ($H_{K2}=1.001H_{K1})$. As can be shown using (\ref
{diag}), for $M_{s}/H_{K1}=0$ , we have $u_{11}=u_{22}=1,$ $v_{11}=v_{22}=0,$
$u_{21}=v_{21}=0,$ $u_{12}=v_{12}=0$. The normal modes are just those of the
individual particles: $(c_{1}=a_{1},c_{2}=a_{2})$ and the damping terms (\ref
{fff}) are just the local damping for the individual grains: 
\begin{eqnarray}
\eta _{1} &=&D(\omega _{1})g^{2}(\omega _{1})  \label{loer} \\
\eta _{2} &=&\beta ^{2}D(\omega _{2})g^{2}(\omega _{2})  \nonumber
\end{eqnarray}

In Fig.3a we plot normalized damping terms $\eta _{1}/D(\omega
_{1})g^{2}(\omega _{1})n(\omega _{1}),\eta _{2}/D(\omega _{2})g^{2}(\omega
_{2})n(\omega _{2})$ versus magnetostatic interaction $M_{s}/H_{K1}$ for the
case of $\beta =1.$ For $M_{s}/H_{K1}=0,$ the damping terms are just those
of the independent grains. As seen in (\ref{gerd}) the damping terms are
proportional to the\ mode frequencies and that variation is seen in Fig.3a.
It is noteworthy that for identical damping mechanisms and a finite grain
interaction,\ the collective modes are not identical.\ 

The case for $\beta =0$, corresponding to relaxation only in one cell, is
shown in Fig.3b. In this example with no interactions ($M_{s}/H_{K1}=0$ and $%
h_{ex}=0),$ $\eta _{2}=0$ corresponds to no thermal bath coupling for grain
2. For a very slight non-uniformity ($H_{K2}=1.001H_{K1}$),\ as $%
M_{s}/H_{K1} $\ increases, the normal modes become a combination of the
individual particles and for $\left( H_{K2}-H_{K1}\right) /H_{K1}=0.001$,
this occurs when the magnetostatic field surpasses the difference of the
anisotropy fields $\left( M_{s}/H_{K1}\approx 0.001\right) .$ For $%
M_{s}/H_{K1}>0.001$ the normalized damping terms become almost equal as the
coupling dominates ($\eta _{1}\approx \eta _{2}\approx 0.5)$ and then
decrease following the frequency dependence in (\ref{gerd}).

In Fig.4 the damping terms are plotted versus exchange for a fixed
magnetostatic interaction $\ M_{s}/H_{K1}=0.09$ and $H_{K2}=1.95H_{K1,}$ as
in Fig.2b. In this case $\beta ^{2}=0.25$, corrsponding to a weaker
relaxation in grain 2 compared to grain 1. The initial values for $h_{ex}=0$
\ reflect the frequency dependence of the damping, between the values as
seen in Figs.3a, 3b. With increasing exchange the damping of mode 1 hardly
varies and the damping of mode 2 increases rapidly, following the results of
Fig.2b and the accompanying discussion.

As a comparison of the results of this analysis with the conventional
LLG-Brown approach, following \cite{bert}, we compare the spectra of thermal
white noise excitation of this system. We take the case of damping only in
mode 1 ($\beta =0).$ The analysis is in Appendix I. For $%
M_{s}/H_{K1}=0.25,H_{K2}=H_{K1,}$ and $h_{ex}=0.5,$ following Fig.4 for the
collective analysis, we use $\eta _{2}/\eta _{1}=1.5$. For the LLG-Brown
equations we use $\alpha _{1}=0.1,$ $\alpha _{2}=0.$ Both curves have been
normalized by matching the first resonance peak and linewidth (by adjusting $%
\eta _{2}/\alpha _{1})$. Both curves exhibit the same resonance frequencies
for the two modes; the resonance frequencies are (to first order)
independent of damping. The spectral shapes differ, however. As in \cite
{bert}, the low frequency PSD of the LLG approach is 3-4 dB above that of
the collective result (greater than 3 dB because of the proximity of the two
resonance frequencies). The second mode peak is lower and broader for the
collective model.

In this work the problem of introducing damping and thermal fluctuations for
an interacting magnetic system is addressed using a physical model of
system-reservoir interactions. No {\it a priori} assumption is made
concerning the form of the dynamic damping. This approach is quite different
from the LLG-Brown approach \cite{brown}, where a dynamic damping is assumed
in the LLG format for each individual grain or discretization cell. In the
LLG-Brown approach, the underlying physical processes of damping and thermal
fluctuations are not explicitly considered. There have been two papers \cite
{smith},\cite{chubykalo} that have attempted to justify the LLG-Brown model
for interacting magnetic units. In \cite{smith} it is argued that the
LLG-Brown approach gives dynamic equations that can be cast in the form of
generalized Newtonian dynamics (specifically\ in the form of an RLC circuit
of coupled oscillators). However, this argument is only inferential and is
not derived using any basic physics model. The result is equivalent to
assuming the application of LLG-Brown to individual grains or discretized
cells of a continuous medium, as is also the essence of \cite{chubykalo}.
These analyses do not derive a stochastic differential equation\ with
damping and thermal fluctuating terms from system-reservoir interactions.

The arguments in \cite{smith},\cite{chubykalo} for independent thermal
fluctuations for independent particles in an interacting magnetic system is,
in fact, a thermodynamic consistent condition required for localized LLG
damping for individual magnetic units. However, in this paper, we have
explicitly constructed a system-reservoir model without any {\it a priori}
assumption of the damping format or thermal fluctuations. The explicit
elimination of thermal reservoir variables in the model gives stochastic
dynamic equations of a damped harmonic oscillator driven by thermal noise
only in the collective normal modes. As we show in Fig.5, different thermal
noise power spectra are obtained. The analysis presented here verifies that
damping and additive fluctuations must be added to the normal modes of an
interacting magnetic system, even if the physical relaxation processes are
local.

\section{Conclusion}

A fundamental analysis of the stochastic dynamic equations for local
coupling to a thermal bath has been performed for a system of two
interacting grains. Here no {\it a priori} assumptions about the form of the
dynamic damping term have been assumed. The results are in the form of
damped harmonic oscillators driven by thermal fields in the collective modes
of the system, a result that previously was derived rigorously for a single
grain with anisotropic energy variations about equilibrium. The LLG-Brown
formalism in which a thermal field is added to each grain or discretization
cell of a continuous medium is shown not to apply. Noise power spectra are
evaluated and it is shown that the two resonance frequencies are broadened
by the collective damping terms. Localized relaxation is felt by the system
collective stochastic dynamics due to the intergranular interactions. The
form of the dynamic damping term, also due to the interactions, reflects the
overall system symmetry. For approximately identical grains these damping
constants are about equal even if the physical damping occurs in only one of
the grains. Noise power spectra are shown which give significant differences
for the two models, thus providing guidance to future experimental analysis.

\begin{acknowledgement}
The authors would like to thank Dr. Vladimir Safonov for numerous helpful
discussions. Support was provided by grant NSF-CMU-EHDR.
\end{acknowledgement}

\section{Appendix I}

The stochastic differential equations (\ref{fff}) for two independent normal
modes can be written as:

\begin{eqnarray}
\frac{dc_{1}}{dt}+\eta _{1}c_{1} &=&-i\omega _{1}c_{1}+f_{1}(t)  \label{ewr}
\\
\frac{dc_{2}}{dt}+\eta _{2}c_{2} &=&-i\omega _{2}c_{2}+f_{2}(t)  \nonumber \\
&<&f_{1}(t)f_{1}^{\ast }(t^{\prime })>=2D_{c1}  \nonumber \\
&<&f_{2}(t)f_{2}^{\ast }(t^{\prime })>=2D_{c2}  \nonumber
\end{eqnarray}
where the ratio of fluctuation magnitudes is: 
\begin{equation}
\frac{D_{c1}}{D_{c2}}=\frac{D(\omega _{1})g^{2}(\omega _{1})n(\omega
_{1})(u_{11}^{2}+\beta ^{2}u_{21}^{2})}{D(\omega _{2})g^{2}(\omega
_{2})n(\omega _{2})(\beta ^{2}u_{22}^{2}+u_{12}^{2})}  \label{ratio}
\end{equation}

For magnetic systems that satisfy a canonical distribution around
equilibrium, the magnitude of damping coefficients $\eta _{1}$ and $\eta
_{2} $ are related to the thermal fluctuations $D_{c1}$ and $D_{c2}$ through
the fluctuation-dissipation condition:

\begin{eqnarray}
\eta _{1} &=&\frac{D_{c1}M_{s}V}{\gamma k_{B}T}\omega _{1}=\alpha
_{Q1}\omega _{1}  \label{fluc-dis} \\
\eta _{2} &=&\frac{D_{c2}M_{s}V}{\gamma k_{B}T}\omega _{2}=\alpha
_{Q2}\omega _{2}  \nonumber
\end{eqnarray}
The spectra of the normal modes can be calculated based upon (\ref{ewr}).
Here the calculation is done in a non-dimensional format and the frequency
is normalized by $\gamma M_{s}$: $\widetilde{\omega }=\omega /\gamma M_{s}$.

The spectral densities for two modes are:

\begin{eqnarray}
S_{1}(\widetilde{\omega }) &=&<c_{1}^{\ast }(\widetilde{\omega })c_{1}(%
\widetilde{\omega })>=\left\vert \frac{1}{(-j\widetilde{\omega }+j\widetilde{%
\omega }_{1}-\alpha _{Q1}\widetilde{\omega }_{1})}\right\vert ^{2}\frac{%
2\alpha _{Q1}k_{B}T}{M_{s}^{2}V}  \label{ccs} \\
S_{2}(\widetilde{\omega }) &=&<c_{2}^{\ast }(\widetilde{\omega })c_{2}(%
\widetilde{\omega })>=\left\vert \frac{1}{(-j\widetilde{\omega }+j\widetilde{%
\omega }_{2}-\alpha _{Q1}\widetilde{\omega }_{2})}\right\vert ^{2}\frac{%
2\alpha _{Q2}k_{B}T}{M_{s}^{2}V}  \nonumber
\end{eqnarray}
The magnetization can be represented by the normal modes as: 
\begin{equation}
\left( 
\begin{array}{c}
m_{1x} \\ 
m_{2x} \\ 
m_{1y} \\ 
m_{2y}
\end{array}
\right) =PU\left( 
\begin{array}{c}
c_{1} \\ 
c_{1}^{\ast } \\ 
c_{2} \\ 
c_{2}^{\ast }
\end{array}
\right)  \label{mtoc}
\end{equation}
where:

\begin{equation}
\text{\ }P=\left( 
\begin{array}{cccc}
1/\sqrt{2} & 1/\sqrt{2} & 0 & 0 \\ 
0 & 0 & 1/\sqrt{2} & 1/\sqrt{2} \\ 
-i/\sqrt{2} & i/\sqrt{2} & 0 & 0 \\ 
0 & 0 & -i/\sqrt{2} & i/\sqrt{2}
\end{array}
\right) \text{\ }  \label{ptran}
\end{equation}
The spectral density function for the magnetization is defined as:

\begin{equation}
S(\omega )=<\overrightarrow{m}(\omega )^{+}\overrightarrow{m}(\omega )>
\label{speLLG}
\end{equation}
where $+$ denotes conjugate transpose and $\overrightarrow{m}%
=[m_{1x},m_{2x,}m_{1y},m_{2y}].$ The magnetization spectral density is
related to the spectral density of the normal modes through the
transformation matrix:

\begin{equation}
TR=U^{+}P^{+}PU  \label{TR}
\end{equation}
Thus, the magnetization spectral density is:

\begin{eqnarray}
S(\widetilde{\omega }) &=&[TR_{11}S_{1}(\widetilde{\omega })+TR_{22}S_{1}(%
\widetilde{\omega })+TR_{33}S_{2}(\widetilde{\omega })+TR_{44}S_{2}(%
\widetilde{\omega })]  \label{seo} \\
&=&\left[ \frac{\alpha _{Q1}(TR_{11}+TR_{22})}{\left| (-j\widetilde{\omega }%
+j\widetilde{\omega }_{1}+\alpha _{Q1}\widetilde{\omega }_{1})\right| ^{2}}+%
\frac{\alpha _{Q2}(TR_{33}+TR_{44})}{\left| (-j\widetilde{\omega }+j%
\widetilde{\omega }_{2}+\alpha _{Q2}\widetilde{\omega }_{1})\right| ^{2}}%
\right] \frac{2k_{B}T}{M_{s}^{2}V}  \nonumber
\end{eqnarray}

For the LLG model, damping and thermal fluctuations are added to individual
particles. For each individual particle:

\begin{equation}
\frac{d\overrightarrow{M}}{dt}=-\gamma \overrightarrow{M}\times 
\overrightarrow{H}-\alpha \gamma \frac{\overrightarrow{M}\times (%
\overrightarrow{M}\times \overrightarrow{H})}{M_{s}}  \label{LLG}
\end{equation}
where $\overrightarrow{H}$ includes the effective field $-\partial
E/\partial \overrightarrow{M}$ and the thermal fluctuating fields $%
\overrightarrow{H}^{T}$. $\ \overrightarrow{H}^{T^{\prime }}$ satisfies the
fluctuation-dissipation condition: $<H_{i}^{T}(t)>=0$ and $%
<H_{i}^{T}(t)H_{j}^{T}(t^{\prime })>=\left( 2\alpha k_{B}T/\gamma
M_{s}V\right) \delta (t-t^{\prime })\delta ij.$ The normalized LLG equations
to leading order are: 
\begin{equation}
\frac{d}{dt}\left( 
\begin{array}{c}
m_{1x} \\ 
m_{2x} \\ 
m_{1y} \\ 
m_{2y}
\end{array}
\right) =\gamma M_{s}A\left( 
\begin{array}{c}
m_{1x} \\ 
m_{2x} \\ 
m_{1y} \\ 
m_{2y}
\end{array}
\right) +\gamma \left( 
\begin{array}{c}
H_{1y}^{T} \\ 
H_{2y}^{^{T}} \\ 
-H_{1x}^{^{T}} \\ 
-H_{1y}^{^{T\prime }}
\end{array}
\right)  \label{nllg}
\end{equation}
where: 
\begin{equation}
\text{\ }A=\left( 
\begin{array}{cccc}
\alpha _{1}(-h_{k1}+1-h_{ex}) & \alpha _{1}(2+h_{ex}) & -h_{k1}+1-h_{ex} & 
-(1-h_{ex}) \\ 
\alpha _{2}(2+h_{ex}) & \alpha _{2}(-h_{k2}+1-h_{ex}) & -(1-h_{ex}) & 
-h_{k2}+1-h_{ex} \\ 
-(-h_{k1}+1-h_{ex}) & -(2+h_{ex}) & \alpha _{1}(-h_{k1}+1-h_{ex}) & -\alpha
_{1}(1-h_{ex}) \\ 
-(2+h_{ex}) & -(-h_{k2}+1-h_{ex}) & -\alpha _{2}(1-h_{ex}) & \alpha
_{2}(-h_{k2}+1-h_{ex})
\end{array}
\right) \text{\ }  \label{GLLG}
\end{equation}
and $\alpha _{1}$ and $\alpha _{2}$ are different damping parameters for the
two particles. The nonzero correlations between fluctuation terms are:

\begin{eqnarray}
&<&H_{1y}^{^{T}}(t)H_{1y}^{^{T}}(t^{\prime })>=\frac{2\alpha _{2}k_{B}T}{%
\gamma M_{s}V}\delta (t-t^{\prime })  \label{noillg} \\
&<&H_{2y}^{^{T}}(t)H_{2y}^{^{T}}(t^{\prime })>=\frac{2\alpha _{2}k_{B}T}{%
\gamma M_{s}V}\delta (t-t^{\prime })  \nonumber \\
&<&H_{1x}^{^{T}}(t)H_{1x}^{^{T}}(t^{\prime })>=\frac{2\alpha _{1}k_{B}T}{%
\gamma M_{s}V}\delta (t-t^{\prime })  \nonumber \\
&<&H_{2x}^{^{T}}(t)H_{2x}^{^{T}}(t^{\prime })>=\frac{2\alpha _{1}k_{B}T}{%
\gamma M_{s}V}\delta (t-t^{\prime })  \nonumber
\end{eqnarray}
Using (\ref{nllg}) , (\ref{GLLG}), we can calculate the nondimesionalized
correlation matrix as:

\begin{equation}
Cor(\widetilde{\omega })=\frac{1}{(j\widetilde{\omega }I-A^{+})}\cdot \frac{1%
}{(-j\widetilde{\omega }I-A)}  \label{cor}
\end{equation}
where $I$ is the unit matrix. The spectral density can be obtained from the
correlation matrix and the fluctuation magnitudes (\ref{noillg}):

\begin{equation}
S(\widetilde{\omega })=[\alpha _{1}Cor_{11}+\alpha _{2}Cor_{22}+\alpha
_{1}Cor_{33}+\alpha _{2}Cor_{44}]\frac{2k_{B}T}{M_{s}^{2}V}  \label{specrl}
\end{equation}


\begin{references}
\bibitem{val0}  V. L. Safonov, {\it J. Mag. Mag. Mate}r., {\bf 195, }523
(1999).

\bibitem{wang0}  X. Wang, N. H. Bertram, and V. L. Safonov, {\it J. Appl.
Phys}. {\bf 92,} 2064 (2002).

\bibitem{bert}  H. N. Bertram, V. L. Safonov, and Z. Jin, {\it IEEE Trans
Magn}., {\bf 38(5)}, pp 2514-2519, Sept. (2002).

\bibitem{LLG}  T . L. Gilbert, Phys. Rev. {\bf 100}, 1243 (1955).

\bibitem{mills}  See the review by D. L. Mills and S. M. Rezende, to be
published in ``Spin Dynamics in Confined Magnetic Structures II'' (Springer
Verlag, Heidelberg), June (2003).

\bibitem{mills2}  D. L. Mills, ``FMR Relaxation in Ultra Thin Films: The
Role of the Conduction Electrons'', submitted to {\it Phys. Rev. B.}

\bibitem{michael}  A. Kunz and R. D. McMichael , {\it IEEE Trans, Magn}., 
{\bf 38} (5), 2400-2402 Part 1 (2002).

\bibitem{brown}  W. F. Brown Jr., {\it Phys. Rev.}, {\bf 130},
1677-1686(1963).

\bibitem{zhu}  J. Zhu, {\it J. Appl. Phys., }{\bf 91}(10), 7273-7275(2002).%
{\it \ }

\bibitem{borner}  E. D. Boerner and H. N. Bertram, {\it IEEE Trans. Magn}., 
{\bf 33} (5), pp. 3052, September 1997.

\bibitem{jin}  Z. Jin, H.N. Bertram and V. L. Safonov, \ {\it IEEE Trans.
Magn}., 38(5), pp. 2265-2267, Sept. 2002.

\bibitem{spark}  M. Sparks, Ferromagnetic Relaxation Theory, McGraw-Hill
(1964).

\bibitem{scully}  M. O. Scully, M. S. Zubairy, ``Quantum Optics'', Cambridge
University Press (1997).

\bibitem{bertzhu}  H. N. Bertram and J-G. Zhu, ``Fundamental Magnetization
Processes in Thin Film Recording Media,''\ in Solid State Physics Review,
Vol. 46, pp. 271-371, eds. H. Ehrenreich and D. Turnbull, Academic Press
(1992).

\bibitem{ReNiFe}  W. Baily, P. Kobes, F. Mancoff, and S. Russek, {\it IEEE
Trans Magn}, {\bf 37}, (4), 1749-54(2001)

\bibitem{Breuer}  H. P. Breuer and F. Petruccione, ``The Theory of Open
Quantum Systems'', Oxford University Press (2002).

\bibitem{xyz}  W. Chen, S. Zhang and N. H. Bertram, J. Appl. Phys. {\bf 71, }%
5579, (1992)

\bibitem{smith}  N. Smith, {\it J. Appl. Phys., }{\bf 92}(7), 3877-3885
(2002).

\bibitem{chubykalo}  O. Chubykalo, U. Nowak, R. Smirnov-Rueda, M.A. Wongsam,
R.W. Chantrell, and J. M. Gonzalez, {\it Phys. Rev. B}, 064422 (2003).
\end{references}
\end{document}